\documentclass{article}
\usepackage{spconf,amsmath,amssymb,graphicx,booktabs,xcolor,tikz,float}
\usepackage{siunitx}
\usetikzlibrary{arrows.meta,calc,fit,positioning}
\usepackage[hidelinks,colorlinks=true]{hyperref}

\definecolor{myblue}{rgb}{0,0.3,0.6}
\hypersetup{linkcolor=myblue,urlcolor=myblue,citecolor=myblue,anchorcolor=myblue}

\newcolumntype{N}{S[table-format=+2.2,detect-weight=true]}
\newcolumntype{P}{S[table-format=2.2,detect-weight=true]}
\newcommand{\CIcells}[3]{#1 & #2 & #3}
\newcommand{\BestNumber}[1]{\multicolumn{1}{r}{\bfseries #1}}
\newcommand{\BestCIcells}[3]{\multicolumn{1}{r@{[}}{\bfseries #1} & \multicolumn{1}{r@{,}}{\bfseries #2} & \multicolumn{1}{r@{]}}{\bfseries #3}}
\newcommand{\NoCIcells}[1]{\multicolumn{1}{N@{}}{#1} & \multicolumn{2}{c@{}}{}}
\newcommand{\BestNoCIcells}[1]{\multicolumn{1}{N@{}}{\bfseries #1} & \multicolumn{2}{c@{}}{}}
\newcommand{\CItext}[3]{\mbox{\num{#1} [\num{#2}, \num{#3}]}}
\newcommand{\BoldCIcells}[3]{\bfseries #1 & \bfseries #2 & \bfseries #3}

\newcommand{\RealRecordingTable}{%
\begin{table*}[!t]
\centering
\ninept
\caption{Museval SDR (dB) on held-out real recordings after SynthSOD and real-recording training. Values are estimates [95\% CI]; the one-piece PHENICX test has no across-recording CI.}
\label{tab:real_recordings}
\setlength{\tabcolsep}{3pt}
\begin{tabular}{@{}lNNNNN@{[}P@{,}P@{]}@{}}
\toprule
System & \multicolumn{1}{c}{Vn.} & \multicolumn{1}{c}{Va.} &
\multicolumn{1}{c}{Vc.} & \multicolumn{1}{c}{Db.} &
\multicolumn{3}{c}{Avg. [95\% CI]} \\
\midrule
\multicolumn{8}{@{}l}{\textbf{URMP} ($n=5$)} \\
HTDemucs & 0.23 & 0.57 & 5.07 & 6.13 & \CIcells{3.00}{0.25}{5.09} \\
BS-RoFormer & 2.46 & 0.86 & 3.31 & 5.32 & \CIcells{2.99}{1.00}{3.51} \\
Tunturi et al., baseline & 1.91 & 1.43 & 5.52 & 5.89 & \CIcells{3.69}{1.63}{4.48} \\
Tunturi et al., score-informed & 2.27 & 1.76 & 5.86 & 5.86 & \CIcells{3.94}{1.66}{5.38} \\
Tunturi et al., score-only & 4.90 & 3.50 & 6.69 & 7.40 & \CIcells{5.62}{4.09}{7.17} \\
SCISSOR (audio only) & 2.21 & 2.30 & 6.64 & 7.65 & \CIcells{4.70}{1.93}{6.60} \\
\textbf{SCISSOR (ours)} & \BestNumber{6.10} & \BestNumber{4.49} & \BestNumber{7.81} & \BestNumber{8.81} & \BestCIcells{6.80}{4.69}{8.98} \\
\midrule
\multicolumn{8}{@{}l}{\textbf{PHENICX-Anechoic} ($n=1$)} \\
HTDemucs & -1.87 & -0.23 & -10.83 & -3.56 & \NoCIcells{-4.12} \\
BS-RoFormer & -2.42 & 0.00 & -1.39 & 5.10 & \NoCIcells{0.32} \\
Tunturi et al., baseline & -0.38 & -2.23 & -2.95 & 3.40 & \NoCIcells{-0.54} \\
Tunturi et al., score-informed & -0.73 & -2.28 & -2.95 & 4.83 & \NoCIcells{-0.29} \\
Tunturi et al., score-only & 0.07 & -4.33 & -4.08 & 1.84 & \NoCIcells{-1.62} \\
SCISSOR (audio only) & 1.00 & -0.09 & \BestNumber{-0.01} & 5.67 & \NoCIcells{1.64} \\
\textbf{SCISSOR (ours)} & \BestNumber{1.28} & \BestNumber{0.10} & -0.02 & \BestNumber{6.00} & \BestNoCIcells{1.84} \\
\bottomrule
\end{tabular}
\end{table*}%
}

\newcommand{\ZeroShotTable}{%
\begin{table*}[!t]
\centering
\ninept
\caption{Avg. museval SDR (dB) after SynthSOD-only training; estimates with 95\% recording-bootstrap CIs.}
\label{tab:zeroshot}
\setlength{\tabcolsep}{2pt}
\begin{tabular}{@{}lcN@{\,[}N@{,}N@{]\hspace{4pt}}N@{\,[}N@{,}N@{]\hspace{4pt}}N@{\,[}N@{,}N@{]}@{}}
\toprule
System & Score & \multicolumn{3}{c}{SynthSOD ($n=59$)} &
\multicolumn{3}{c}{PHENICX ($n=4$)} &
\multicolumn{3}{c}{URMP ($n=22$)} \\
\midrule
Mixture (no separation) & & \CIcells{-7.45}{-8.69}{-6.36} & \CIcells{-14.11}{-17.80}{-8.26} & \CIcells{-5.23}{-5.72}{-3.38} \\
HTDemucs~\cite{htdemucs} & & \CIcells{4.17}{2.43}{5.20} & \CIcells{-1.83}{-5.74}{1.20} & \CIcells{1.93}{0.75}{3.31} \\
BS-RoFormer~\cite{bsroformer} & & \CIcells{3.39}{2.30}{4.18} & \CIcells{-2.72}{-5.33}{0.16} & \CIcells{1.54}{0.84}{2.63} \\
Tunturi et al., baseline~\cite{scoremss} & & \CIcells{5.64}{4.23}{6.27} & \CIcells{-0.15}{-1.51}{2.35} & \CIcells{1.88}{0.99}{2.69} \\
Tunturi et al., score-informed & \checkmark & \CIcells{5.96}{4.59}{6.53} & \CIcells{0.55}{-0.69}{2.35} & \CIcells{3.35}{2.17}{3.94} \\
Tunturi et al., score-only & \checkmark & \CIcells{4.35}{3.40}{5.01} & \CIcells{0.83}{-2.43}{3.39} & \BoldCIcells{4.80}{4.13}{5.96} \\
\midrule
SCISSOR (audio only) & & \CIcells{5.97}{4.74}{6.56} & \CIcells{1.43}{0.18}{2.50} & \CIcells{1.88}{1.21}{2.47} \\
\textbf{SCISSOR (ours)} & \checkmark & \BoldCIcells{6.31}{5.15}{6.90} & \BoldCIcells{1.85}{1.16}{2.78} & \CIcells{2.52}{1.86}{3.34} \\
\bottomrule
\end{tabular}
\end{table*}%
}

\newcommand{\CorruptionTable}{%
\begin{table*}[!t]
\centering
\ninept
\caption{Whole-piece SDR (dB) under score corruption on 22 SynthSOD test pieces; estimates with 95\% piece-bootstrap CIs.}
\label{tab:corruption_synth}
\setlength{\tabcolsep}{4pt}
\begin{tabular}{@{}lrrr@{}}
\toprule
Score error & Tunturi score-informed~\cite{scoremss} & Tunturi score-only & SCISSOR \\
\midrule
Clean score & \CItext{4.81}{4.50}{5.11} & \CItext{2.92}{2.49}{3.33} & \CItext{5.45}{5.13}{5.76} \\
Matched audio-only control & \CItext{4.58}{4.29}{4.89} & \CItext{4.58}{4.29}{4.89} & \CItext{4.98}{4.67}{5.29} \\
\midrule
Onset jitter ($\sigma=400$\,ms) & \CItext{4.38}{4.10}{4.67} & \CItext{2.15}{1.79}{2.50} & \CItext{5.23}{4.91}{5.53} \\
Global offset ($1$\,s) & \CItext{3.92}{3.64}{4.21} & \CItext{1.78}{1.45}{2.10} & \CItext{4.97}{4.67}{5.27} \\
Tempo scaling ($10\%$) & \CItext{3.35}{3.10}{3.61} & \CItext{0.99}{0.75}{1.22} & \CItext{4.72}{4.42}{5.01} \\
Pitch substitution ($50\%$) & \CItext{4.34}{4.04}{4.63} & \CItext{2.10}{1.70}{2.46} & \CItext{5.29}{4.98}{5.59} \\
Note deletion ($75\%$) & \CItext{3.40}{3.14}{3.66} & \CItext{0.73}{0.53}{0.92} & \CItext{5.04}{4.72}{5.35} \\
Missing parts (two) & \CItext{3.77}{3.44}{4.11} & \CItext{1.11}{0.62}{1.60} & \CItext{5.19}{4.88}{5.49} \\
Instrument relabelling ($100\%$) & \CItext{2.70}{2.42}{2.98} & \CItext{-0.58}{-0.97}{-0.26} & \CItext{3.92}{3.65}{4.19} \\
Part swaps (both pairs) & \CItext{2.14}{1.73}{2.49} & \CItext{-0.60}{-1.05}{-0.21} & \CItext{3.45}{3.12}{3.75} \\
Structural cut ($20\%$) & \CItext{3.75}{3.46}{4.06} & \CItext{1.35}{1.06}{1.68} & \CItext{5.00}{4.66}{5.32} \\
Combined, severe & \CItext{4.05}{3.78}{4.32} & \CItext{1.71}{1.40}{2.01} & \CItext{5.04}{4.73}{5.33} \\
Score withheld & \CItext{2.87}{2.60}{3.15} & \CItext{0.09}{0.06}{0.12} & \CItext{4.96}{4.64}{5.26} \\
\bottomrule
\end{tabular}
\end{table*}%
}

\title{SCISSOR: SCORE-CONDITIONED INSTRUMENT SOURCE SEPARATION FOR ORCHESTRAL RECORDINGS}
\name{
  Yiheng Lu$^{1}$ \qquad
  Hao-Wen Dong$^{2}$\setcounter{footnote}{1}\sthanks{Corresponding author: hwdong@umich.edu}
}
\address{
  $^1$The Chinese University of Hong Kong, Shenzhen\qquad
  $^2$University of Michigan
}

\begin{document}
\ninept
\maketitle

\begin{abstract}
Orchestral separation recovers instrument sections from mixtures in which
shared pitches, harmonics, and timbres obscure source identity. An aligned
score provides instrument labels, note pitches, and activity times. A
score-informed approach appends piano rolls to audio features before mask
prediction. We
introduce \emph{SCISSOR} (Score-Conditioned Instrument Source Separation for
Orchestral Recordings), which uses the score to form a frame-wise query for
each source. Each query matches a shared audio representation, and a softmax
over instrument and background slots jointly assigns overlapping
time--frequency evidence. The queries retain instrument identity even when
notes are missing from the score. After training on SynthSOD and a small
set of URMP and PHENICX-Anechoic recordings, SCISSOR achieves the highest
average SDR on held-out real recordings. With SynthSOD-only training, it
leads on SynthSOD and zero-shot PHENICX-Anechoic, and improves on its
audio-only control on zero-shot URMP. SCISSOR also degrades less under score
corruption than the evaluated score-based baselines.
\end{abstract}


\section{Introduction}
\label{sec:intro}

Music source separation (MSS) recovers individual sources from a music
recording~\cite{rafii2018overview}. In orchestral MSS, these sources are
instrument sections, rather than the vocal, bass, drum, and accompaniment
stems often studied in popular music~\cite{synthsod}. Sections may play the
same pitch and produce similar timbres and overlapping harmonics. As a
result, the mixture alone may not reveal which section produced a given
sound~\cite{xu2025position}.

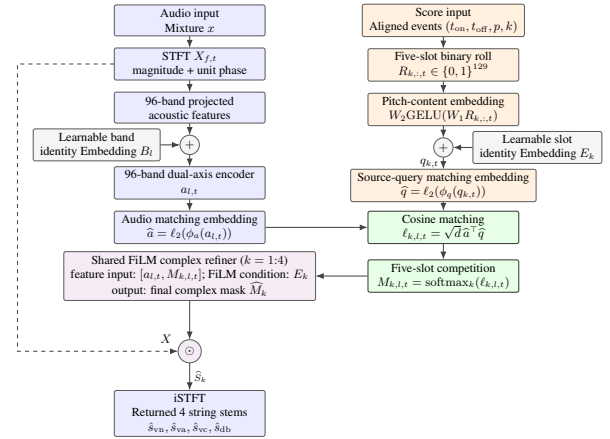
\begin{figure}
\centering
\resizebox{0.90\columnwidth}{!}{%
\begin{tikzpicture}[
  font=\small,
  >=Latex,
  node distance=3mm and 24mm,
  box/.style={
    draw=black!65, rounded corners=1.5pt, fill=black!4,
    align=center, minimum height=8mm, minimum width=39mm,
    inner xsep=5pt, inner ysep=2.5pt
  },
  audio/.style={box, fill=blue!8},
  score/.style={box, fill=orange!12},
  identity/.style={box, minimum width=24mm, minimum height=7mm},
  add/.style={circle, draw=black!65, fill=black!4, inner sep=1pt,
    minimum size=5mm},
  product/.style={circle, draw=black!65, fill=violet!8, inner sep=1pt,
    minimum size=6mm},
  fusion/.style={box, fill=green!10, draw=black!80, line width=.6pt},
  output/.style={box, fill=violet!8},
  flow/.style={->, line width=.6pt, draw=black!75}
]

\node[audio] (wav) {Audio input\\Mixture $x$};
\node[audio, below=of wav] (tf)
  {STFT $X_{f,t}$\\magnitude + unit phase};
\node[audio, below=of tf] (band)
  {96-band projected\\acoustic features};
\node[add, below=of band] (bsum) {$+$};
\node[audio, identity, left=5mm of bsum] (bid)
  {Learnable band\\identity Embedding $B_l$};
\node[audio, below=of bsum] (encoder)
  {96-band dual-axis encoder\\$a_{l,t}$};
\node[audio, below=of encoder] (akey)
  {Audio matching embedding\\$\widehat a=\ell_2\!\left(\phi_a(a_{l,t})\right)$};

\node[score, right=25mm of wav] (events)
  {Score input\\Aligned events $(t_{\rm on},t_{\rm off},p,k)$};
\node[score, below=of events] (roll)
  {Five-slot binary roll\\$R_{k,:,t}\in\{0,1\}^{129}$};
\node[score, below=of roll] (pitch)
  {Pitch-content embedding\\$W_2\mathrm{GELU}(W_1R_{k,:,t})$};
\node[add, below=of pitch] (esum) {$+$};
\node[score, identity, right=5mm of esum] (eid)
  {Learnable slot\\identity Embedding $E_k$};
\node[score, below=of esum] (qkey)
  {Source-query matching embedding\\$\widehat q=\ell_2\!\left(\phi_q(q_{k,t})\right)$};

\draw[flow] (wav) -- (tf);
\draw[flow] (tf) -- (band);
\draw[flow] (band) -- (bsum);
\draw[flow] (bid) -- (bsum);
\draw[flow] (bsum) -- (encoder);
\draw[flow] (encoder) -- (akey);
\draw[flow] (events) -- (roll);
\draw[flow] (roll) -- (pitch);
\draw[flow] (pitch) -- (esum);
\draw[flow] (eid) -- (esum);
\draw[flow] (esum) -- node[left, midway] {$q_{k,t}$} (qkey);

\node[fusion] (match) at (qkey.center |- akey.center)
  {Cosine matching\\$\ell_{k,l,t}=\sqrt d\,\widehat a^{\top}\widehat q$};
\node[fusion, below=4mm of match] (softmax)
  {Five-slot competition\\$M_{k,l,t}=\operatorname{softmax}_k(\ell_{k,l,t})$};

\draw[flow] (akey.east) -- (match.west);
\draw[flow] (qkey.south) -- (match.north);
\draw[flow] (match) -- (softmax);

\node[output] (refiner) at (akey.center |- softmax.center)
  {Shared FiLM complex refiner ($k=1{:}4$)\\
  feature input: $[a_{l,t},M_{k,l,t}]$; FiLM condition: $E_k$\\
  output: final complex mask $\widehat M_k$};
\node[product, below=10mm of refiner] (multiply) {$\odot$};
\node[audio, below=7mm of multiply] (stems)
  {iSTFT\\Returned 4 string stems\\
  $\hat s_{\rm vn},\hat s_{\rm va},\hat s_{\rm vc},\hat s_{\rm db}$};

\draw[flow] (softmax.west) -- (refiner.east);
\draw[flow] (refiner) -- (multiply);
\coordinate (mixroute) at ($(bid.west)+(-4mm,0)$);
\draw[flow, dashed] (tf.west) -- (mixroute |- tf.west)
  -- (mixroute |- multiply.west) -- (multiply.west);
\node[above=1mm of multiply.west, xshift=-3mm] {$X$};
\draw[flow] (multiply) -- node[right, midway, font=\scriptsize] {$\widehat S_k$} (stems);

\end{tikzpicture}%
}
\caption{SCISSOR overview: score-derived source queries compete for shared audio
evidence, and a shared refiner reconstructs four string stems.}
\label{fig:architecture}
\end{figure}

Before deep learning, audio-only MSS commonly used nonnegative matrix
factorization (NMF) to decompose mixture spectrograms, with constraints on
temporal continuity and sparsity~\cite{virtanen2007nmf}. Neural separators
learn source estimates from data. Spectrogram-based models predict masks
with convolutional networks~\cite{jansson2017unet}, recurrent networks such
as X-UMX~\cite{xumx}, or subband models such as BSRNN and
BS-RoFormer~\cite{bsrnn,bsroformer}. HTDemucs combines spectral and waveform
representations~\cite{htdemucs}. Despite these advances, audio-only models
must infer instrument activity and identity from the mixture itself, which
is difficult when orchestral sources overlap in pitch and timbre.

An aligned score specifies each instrument's note pitches and onset and
offset times. Earlier score-informed methods used these cues to constrain
NMF~\cite{ewert2012nmf,fritsch2013nmf} or to guide neural training with
score-filtered spectra and weak labels~\cite{miron2017,ewert2017}. More
recently, Tunturi et al.~\cite{scoremss} encoded instrument piano rolls and
concatenated them with the mixture spectrogram in an X-UMX separator. They
also tested a score-only variant that predicts masks without audio
features. Score concatenation helped on small ensembles, and the score-only
variant transferred well to some real recordings. In denser orchestral
mixtures, however, the gains from concatenation were less consistent, and
the score-informed model struggled to transfer from synthetic to real
audio. The effect of score misalignment was not examined~\cite{scoremss}.

We propose \emph{SCISSOR}, a score-conditioned source-query method for
orchestral separation. A shared encoder represents the mixture across time
and frequency. At each frame, SCISSOR combines active score pitches with a
learned instrument identity to form a query for each source. Even if the
score omits a note, the identity remains in the query and keeps that source
available. Separate projections map audio features and score queries into
a shared space, where their normalized embeddings are matched at each
band--frame location. A softmax over four instrument slots and a background
slot then assigns mixture evidence jointly: a stronger match for one source
leaves less evidence for the others. A shared complex refiner uses these
soft assignments and the audio features to reconstruct the instrument
stems. Joint query competition lets acoustic evidence temper
inaccurate score cues.

\noindent\textbf{Our contributions are as follows:}
\begin{itemize}
  \item \textbf{A new score-informed separation method.} SCISSOR uses
  frame-wise instrument queries to match score cues with encoded audio and
  allocate mixture evidence jointly across sources.
  \item \textbf{Improved separation quality.} SCISSOR achieves the highest
  average SDR among the evaluated systems on held-out real recordings and
  SynthSOD.
  \item \textbf{Effective use of limited real data and robust score guidance.}
  SCISSOR performs well with a small real-recording training set and loses
  less SDR under score corruption than the evaluated score-based baselines.
\end{itemize}

We train SCISSOR on aligned scores and audio from synthetic SynthSOD and
real URMP and PHENICX-Anechoic recordings~\cite{synthsod,urmp,phenicx}.
We evaluate separation on held-out synthetic and real recordings, zero-shot
transfer to real audio, and robustness to score corruption.

Audio examples are available at
\url{https://scissorsep.github.io/}.
All project code and model checkpoints will be made publicly available upon
acceptance.

\section{Method}
\label{sec:method}
SCISSOR extracts four string sections from a full-orchestra mixture, as shown
in Fig.~\ref{fig:architecture}. Given a mono mixture $x$ and an aligned
symbolic score $\mathcal{S}$, the model estimates violin, viola, cello, and
double-bass signals ($K=4$). The audio pathway computes the log magnitude
and normalized phase of the mixture spectrum $X$, projects these features
onto $L=96$ overlapping quasi-log-frequency bands, and adds a learned band
identity. Dual-axis attention then produces an audio representation
$a_{l,t}$ for band $l$ and frame $t$. The score pathway maps note events to
four target slots and one background slot, rasterizes pitch activity on the
audio frame grid, and combines each activity vector with a learned slot
identity. Separate audio and score projections yield scaled cosine
compatibilities; a softmax over the five slots jointly allocates each
band--frame region. A shared identity-conditioned complex refiner uses the
four target allocations and audio features to estimate complex masks
$\widehat M_k$ and sources $\widehat S_k=\widehat M_k\odot X$. The background
allocation absorbs non-target energy internally. The following components
specify the query construction, matching, and source assignment.

\RealRecordingTable

\noindent\textbf{Frame-wise source queries.}\par\nobreak\noindent Each score event specifies an onset, offset, pitch, and MIDI program.
Events for the four requested string sections enter their respective target
slots; other programs enter the background slot. We dilate onsets by 30~ms
and offsets by 80~ms, then rasterize the events on the audio frame grid as
$R\in\{0,1\}^{K'\times129\times T}$, where $K'=K+1$. A source query combines
the active pitches of slot $k$ with its learned identity $E_k$:
\begin{equation}
q_{k,t}=E_k+W_2\,\mathrm{GELU}\!\left(W_1R_{k,:,t}\right).
\label{eq:query}
\end{equation}
The vector $R_{k,:,t}\in\{0,1\}^{129}$ records the pitches of source $k$
at frame $t$; $W_1$ and $W_2$ embed that activity, and
$E_k\in\mathbb{R}^{d}$ identifies the source. Thus $q_{k,t}$ represents both
the expected instrument and its active pitches. Because the two linear layers
have no bias, an empty pitch vector yields $q_{k,t}=E_k$. The instrument
therefore remains a candidate when the score omits a note or misaligns its
timing.

\noindent\textbf{Cross-modal matching embeddings.}\par\nobreak\noindent SCISSOR projects audio features and score queries separately into a
shared matching space:
\begin{equation}
\begin{aligned}
\phi_r(z)&=V_r\,\mathrm{GELU}\!\left(U_r\mathrm{LN}(z)+b_r\right),
\quad r\in\{a,q\},\\
\widehat a_{l,t}&=\ell_2\!\left(\phi_a(a_{l,t})\right),\qquad
\widehat q_{k,t}=\ell_2\!\left(\phi_q(q_{k,t})\right).
\end{aligned}
\label{eq:projections}
\end{equation}
The index $r\in\{a,q\}$ selects the audio or query projection, and
$\ell_2(v)=v/\lVert v\rVert_2$. Separate projections accommodate the input
statistics of the two modalities. Unit normalization makes
$\widehat a_{l,t}^{\top}\widehat q_{k,t}$ a cosine compatibility between
the shared audio representation and a source query.

\noindent\textbf{Score-conditioned instrument source separation.}\par\nobreak\noindent The $K'=K+1$ queries comprise four requested string sections and a
background slot containing non-target score programs. For each audio
band--frame representation, SCISSOR computes query compatibilities and
normalizes them over the source axis:
\begin{equation}
M_{k,l,t}=\mathrm{softmax}_{k}\!\left(
\sqrt{d}\,\widehat a_{l,t}^{\top}\widehat q_{k,t}\right).
\label{eq:match}
\end{equation}
The fixed factor $\sqrt{d}$ scales cosine compatibility, and $M_{k,l,t}$
denotes the allocation to source $k$ at band $l$ and frame $t$. The softmax
enforces $\sum_{k=1}^{K'}M_{k,l,t}=1$: a larger allocation to one source
reduces the allocation available to its competitors. The background slot
absorbs non-target energy without producing an output stem. Because the
allocations remain soft, acoustic evidence can counter an absent or
incorrect score event.

\noindent\textbf{Training and implementation details.}\par\nobreak\noindent We train SCISSOR end to end with
\begin{equation}
\mathcal{L}=\mathcal{L}_{\mathrm{wav}}+\mathcal{L}_{\mathrm{cSTFT}}
+0.25\mathcal{L}_{\mathrm{mask}}+0.15\mathcal{L}_{\mathrm{comp}}
+0.15\mathcal{L}_{\mathrm{SI}},
\label{eq:loss}
\end{equation}
The five terms measure waveform $\ell_1$, complex-STFT $\ell_1$,
band-ratio-mask error, competition-region error, and scale-invariant waveform
error, respectively. A small loss weight on silent targets discourages
leakage. For each requested source, a shared FiLM-conditioned refiner takes
$[a_{l,t},M_{k,l,t}]$, predicts real and imaginary band-domain corrections,
and converts them to a bounded complex mask. The model uses $d=448$, a
256-dimensional refiner state, and nine dual-axis audio blocks. Unless
specified otherwise, experiments use 44.1-kHz audio, a 4096-point STFT with
hop size 1024, 96 bands, and the score augmentations in Section~\ref{sec:exp}.

\section{Experiment}
\label{sec:exp}

\noindent\textbf{SynthSOD.}\par\nobreak\noindent SynthSOD~\cite{synthsod,scoremss} contains synthesized ensembles and
orchestras with isolated stems spanning 18 instrument tracks: strings,
woodwinds, brass, harp, and percussion. Our 484-piece subset has two to
14 active tracks per piece and aligned score events specifying each note's
onset, offset, pitch, and instrument. We use 351 pieces for training, 42 for
validation, and 91 for testing. The input is the mono full mixture from the
Tree microphone. Violin, viola, cello, and double bass are the targets; all
other stems remain in the mixture. Among the test pieces, 59 contain at
least one target instrument.

\noindent\textbf{URMP.}\par\nobreak\noindent URMP~\cite{urmp} contains 44 real chamber-music recordings with two to
five parts per piece, covering 13 instrument types across strings,
woodwinds, and brass. Its score and note annotations are aligned with the
recorded parts; we use note onset, offset, pitch, and instrument labels.
We select the 22 recordings with string targets and split them by musical
work into 13 training, four validation, and five test pieces. All four
requested string instruments are evaluated when present.

\noindent\textbf{PHENICX-Anechoic.}\par\nobreak\noindent PHENICX-Anechoic~\cite{phenicx} contains four real orchestral excerpts
with 10 to 39 separately recorded parts per piece. They cover up to ten
instrument types across strings, woodwinds, and brass. The aligned score
provides note onset, offset, pitch, and instrument labels, alongside
isolated audio references. We use two pieces for training, one for
validation, and one for testing. Across URMP and PHENICX-Anechoic, the
six test pieces come from three works absent from training and validation.

\noindent\textbf{Implementation details.}\par\nobreak\noindent SCISSOR processes 44.1-kHz mono audio with a 4096-point STFT and a
1024-sample hop. Log-magnitude and unit-phase features are projected onto
96 frequency bands; audio and score queries use 448-dimensional embeddings.
SCISSOR has 28.04 million parameters. We also train an audio-only SCISSOR
with the same architecture and parameter count. It receives an empty score
during training, validation, and inference, leaving only the learned
instrument identities in its source queries; its data, loss, optimizer, and
training budget match those of SCISSOR.
We train on four NVIDIA RTX 4080 GPUs. SynthSOD training runs from random
initialization for 20,000 optimizer steps on 4-s crops with global batch
size 12. We use AdamW with learning rate $1.5\times10^{-4}$, 1,500 warm-up
steps, cosine decay, and an exponential moving average of 0.999. Real
recording training starts from the SynthSOD checkpoint and runs for 2,000
steps with effective batch size 12. During training, we randomly withhold
the score, jitter note onsets, and drop pitch cells.

\noindent\textbf{Evaluation metrics.}\par\nobreak\noindent We measure separation quality with
signal-to-distortion ratio (SDR, in dB); higher values indicate better
separation. Tables~\ref{tab:real_recordings} and~\ref{tab:zeroshot} use
museval BSSEval~v4 SDR~\cite{sisec2018}, computed in one-second windows. For each instrument,
we take the median across windows within each recording and then the median
across recordings. \emph{Avg.} is the unweighted mean of the four instrument
scores; silent references are excluded. Table~\ref{tab:corruption_synth}
instead uses whole-piece energy-ratio SDR, averaged over active instruments
within each piece.

For aggregates over at least two recordings or pieces, we report 95\%
confidence intervals (CIs) from the 2.5th and 97.5th percentiles of 10,000
bootstrap replicates (seed 20260924). We resample whole recordings or pieces
with replacement and
recompute the target-median macro-average for
Tables~\ref{tab:real_recordings} and~\ref{tab:zeroshot}, or the mean of
piece-level target averages for Table~\ref{tab:corruption_synth}. One-second
windows are not treated as independent samples. Results are shown as
\emph{estimate} [95\% CI]; the single held-out PHENICX-Anechoic recording
has no across-recording CI.

\noindent\textbf{Baseline models and comparison setup.}\par\nobreak\noindent We compare SCISSOR with HTDemucs~\cite{htdemucs} (41.98M parameters),
BS-RoFormer~\cite{bsroformer} (58.07M), and the audio-only (28.32M),
score-informed (29.77M), and score-only (25.50M) X-UMX models of
Tunturi et al.~\cite{scoremss}, as well as the matched audio-only SCISSOR
(28.04M).
All systems use the aligned SynthSOD training and validation sets. For
Table~\ref{tab:real_recordings}, they then train on the same aligned URMP
and PHENICX-Anechoic recordings. Audio-only models receive audio
without score cues. HTDemucs and BS-RoFormer use the same SynthSOD 351/42
split and training budget as SCISSOR, while the X-UMX models start from
released SynthSOD checkpoints. We use common mixtures and scoring code and
evaluate only the violin, viola, cello, and double-bass outputs.

\vspace{1ex}
\ZeroShotTable
\CorruptionTable

\noindent\textbf{Real-recording separation test setup (Table~\ref{tab:real_recordings}).}\par\nobreak\noindent Every model is first trained on SynthSOD, then trained on 15 aligned
real recordings (13 URMP and two PHENICX-Anechoic), with five others
(four URMP and one PHENICX-Anechoic) used for model selection. We then
evaluate separation on five held-out URMP recordings and one held-out
PHENICX-Anechoic recording, reporting the two corpora separately.

\noindent\textbf{Synthetic and zero-shot separation test setup (Table~\ref{tab:zeroshot}).}\par\nobreak\noindent Using only SynthSOD-trained checkpoints, we compare the same systems
on the 59 test pieces with string targets. We then test transfer to all 22
URMP and four PHENICX-Anechoic recordings without real-recording training.
The unseparated mixture provides a performance floor.

\noindent\textbf{Score-corruption test setup (Table~\ref{tab:corruption_synth}).}\par\nobreak\noindent We compare SCISSOR with Tunturi et al.'s score-informed and score-only
models on 22 SynthSOD test pieces containing all four strings. Across 33
conditions, we perturb note timing, pitch, presence, instrument identity,
and score structure, including combined errors. The displayed severe
conditions include 400-ms onset jitter, 50\% pitch substitution, 75\%
note deletion, and complete instrument relabelling. For each condition,
the checkpoint, mixture, references, and inference procedure remain fixed;
only the score changes. Clean scores, withheld scores, and matched audio-only
models serve as controls.

\section{Results}
\label{sec:results}

\noindent\textbf{Real-recording separation test results.}\par\nobreak\noindent Table~\ref{tab:real_recordings} compares systems trained on SynthSOD and
real recordings. On five held-out URMP pieces, SCISSOR reaches 6.80~dB
Avg. SDR [4.69, 8.98], a 1.18-dB gain over the strongest external baseline,
Tunturi et al.'s score-only model at 5.62~dB [4.09, 7.17]. Relative to the
best external baseline for each instrument, the gains are 1.20, 0.99, 1.12,
and 1.41~dB for violin, viola, cello, and double bass, respectively. On
the held-out PHENICX-Anechoic recording, SCISSOR improves Avg. SDR by
1.52~dB over the best external baseline (1.84 versus 0.32~dB); the corresponding
instrument gains are 1.21, 0.10, 1.37, and 0.90~dB. Its cello result is
0.01~dB below its own audio-only control. The URMP confidence intervals
overlap, so these gains do not establish statistical significance. The
single PHENICX test piece has no across-recording confidence interval.

\noindent\textbf{Synthetic and zero-shot separation test results.}\par\nobreak\noindent Table~\ref{tab:zeroshot} uses checkpoints trained only on SynthSOD. On
the 59-piece synthetic test, SCISSOR has the highest Avg. SDR at 6.31~dB
[5.15, 6.90], compared with 5.96~dB [4.59, 6.53] for the strongest
external baseline. It also has the highest point estimate on four zero-shot
PHENICX recordings, 1.85~dB [1.16, 2.78], although its interval overlaps
that of its audio-only control at 1.43~dB [0.18, 2.50]. On 22 zero-shot
URMP recordings, SCISSOR improves on its control (2.52 versus 1.88~dB),
but Tunturi et al.'s score-only model remains higher at 4.80~dB
[4.13, 5.96]. The SynthSOD intervals also overlap, whereas the URMP
intervals for SCISSOR [1.86, 3.34] and the score-only model do not. Thus,
the synthetic-data advantage does not extend to every recorded ensemble.

\noindent\textbf{Score-corruption test results.}\par\nobreak\noindent \label{sec:corruption} Table~\ref{tab:corruption_synth} shows representative errors from 33 score
corruptions. With 75\% of notes deleted, SCISSOR loses 0.41~dB relative to
its clean score, compared with losses of 1.41~dB for Tunturi et al.'s
score-informed model and 2.19~dB for its score-only model. The resulting
SDRs are 5.04 [4.72, 5.35], 3.40 [3.14, 3.66], and 0.73~dB
[0.53, 0.92], with non-overlapping 95\% CIs. Complete instrument
relabelling causes larger drops of 1.53, 2.11, and 3.50~dB, respectively;
SCISSOR still has the highest SDR at 3.92~dB [3.65, 4.19], compared with
2.70~dB [2.42, 2.98] and $-0.58$~dB [$-0.97$, $-0.26$]. Across all 33
conditions, SCISSOR falls below its audio-only control in 8, versus 23
for score-informed and 33 for score-only separation. Instrument-identity
errors remain challenging: relabelling and part swaps reduce SCISSOR below
its 4.98-dB audio-only control.

\section{Conclusion}
\label{sec:conclusion}

SCISSOR forms frame-wise source queries from score pitches and instrument
identities, matches them to shared audio features, and jointly assigns
mixture energy to instrument and background slots. Identity-only queries
keep instruments available when notes are omitted. With a small
real-recording training set, SCISSOR led the evaluated systems on held-out
URMP and PHENICX-Anechoic recordings. Trained only on SynthSOD, it led on
SynthSOD and zero-shot PHENICX-Anechoic and surpassed its audio-only
control on zero-shot URMP; Tunturi et al.'s score-only model led on the
latter. SCISSOR degraded less under representative score errors, although
instrument-identity errors remain challenging.

\begin{center}\textbf{ACKNOWLEDGMENTS}\end{center}
No funding was received for this work. The authors declare no relevant
financial or nonfinancial conflicts of interest.

\begin{center}\textbf{COMPLIANCE WITH ETHICAL STANDARDS}\end{center}
This study used only publicly available music datasets and involved no new
human or animal subject experiments; no ethics approval was required.

\bibliographystyle{IEEEbib}
{\bibliography{strings,refs}}

\end{document}